\documentclass[twocolumn,showpacs,preprintnumbers,amsmath,amssymb]{revtex4}

\usepackage{graphicx}
\usepackage{dcolumn}
\usepackage{bm}

\newcommand{\be}{\begin{equation}}
\newcommand{\ee}{\end{equation}}
\newcommand{\bea}{\begin{eqnarray}}
\newcommand{\eea}{\end{eqnarray}}
\newcommand{\ba}{\begin{eqnarray}}
\newcommand{\ea}{\end{eqnarray}}
\newcommand{\nn}{\nonumber \\}
\newcommand{\eqn}[1]{(\ref{#1})}
\newcommand{\beq}{\begin{equation}}
\newcommand{\eeq}{\end{equation}}
\newcommand{\beqa}{\begin{eqnarray}}
\newcommand{\eeqa}{\end{eqnarray}}
\newcommand{\beqar}{\begin{eqnarray*}}
\newcommand{\eeqar}{\end{eqnarray*}}

\newcommand{\ie}{{\it i.e.,}\ }

\def\sac{\, , \,\,\,\,\,}

\def\nc {N_\mt{c}}
\def\nf {N_\mt{f}}

\def\t6 {T_\mt{D6}}

\def\gym {g_\mt{YM}}

\newcommand{\te}{t_\mt{E}}
\newcommand{\td}{T_\mt{deconf}}
\newcommand{\tf}{T_\mt{fund}}
\newcommand{\leff}{g_\mt{eff}}

\newcommand{\mq}{M_\mt{q}}      
\newcommand{\mqs}{M_\mt{q}^*}      
\newcommand{\qc}{\langle \bar{\psi} \psi \rangle} 
\newcommand{\qcs}{\qc^*} 
\newcommand{\tq}{T_\mt{Dq}}      
\newcommand{\R}{L} 
\newcommand{\Y}{y} 
\newcommand{\Z}{z} 
\newcommand{\kk}{\mu} 
\newcommand{\N}{{\cal N}} 
\newcommand{\mbar}{\bar{M}}
\newcommand{\rhomin}{\rho_\mt{min}}
\newcommand{\rhomax}{\rho_\mt{max}}
\newcommand\ict{I_{\mt{bound}}}

\newcommand{\ibk}{I_\mt{Dq}}

\newcommand{\mt}[1]{\textrm{\tiny #1}}

\begin{document}

\preprint{hep-th/0605046}

\title{Holographic Phase Transitions with Fundamental Matter}

\author{David Mateos,$^1$ Robert C. Myers,$^{2,3}$ and
Rowan M. Thomson,$^{2,3}$}
\affiliation{$^1\,$Department of Physics, University of California, Santa
Barbara, CA 93106-9530, USA \\
$^2 \,$Perimeter Institute for Theoretical Physics, Waterloo,
Ontario N2J 2Y5, Canada \\
$^3 \,$Department of Physics, University of Waterloo,
Waterloo, Ontario N2L 3G1, Canada
}


\begin{abstract}
The holographic dual of a finite-temperature gauge theory with a
small number of flavours typically contains D-brane probes in a
black hole background. At low temperature  the branes sit outside
the black hole and the meson spectrum is discrete and
possesses a mass gap. As the
temperature increases the branes approach a critical solution.
Eventually they fall into the horizon and a phase transition
occurs. In the new phase the meson spectrum is continuous and
gapless. At large $\nc $ and large 't Hooft coupling, this phase
transition is always of first order, and in confining theories with
heavy quarks it occurs at a temperature higher than the deconfinement
temperature for the glue.
\end{abstract}

\maketitle

\noindent {\bf Introduction:} The gauge/gravity correspondence is a
powerful tool to study non-perturbative physics of gauge theories in
diverse dimensions. The classical supergravity regime corresponds to
the large-$\nc$, strong 't Hooft coupling limit of the gauge theory.
This allows the study of a large class of theories that share some
of the important features of four-dimensional QCD, such as
confinement, chiral symmetry breaking, thermal phase transitions,
etc. In principle, because of its asymptotic freedom, QCD itself is
not in this class. This means that calculations of certain
quantitative properties of QCD, such as the detailed mass spectrum,
for example, will require going beyond the supergravity
approximation. However, this does not exclude the possibility that
some aspects of QCD can be studied in this approximation: Some
predictions of the gauge/gravity correspondence may be universal
enough as to apply to QCD, at least in certain regimes. A suggestive
recent example is the gauge/gravity calculation of the shear
viscosity in the hydrodynamic regime of strongly coupled
finite-temperature gauge theories. The viscosity/entropy ratio is
universal for a large class of gauge theories in the regime
described by their gravity duals  \cite{hydro} which, for high
enough a temperature, generically contain a black hole horizon.  Moreover, this ratio appears to be surprisingly close to that inferred from
experiments at the Relativistic Heavy Ion Collider (RHIC)
\cite{rhic}. It is therefore important to establish as many
universal features of the gauge/gravity correspondence as possible.

The quarks in QCD transform in the fundamental representation. For a
large class of gauge theories,  a small number of flavours of
fundamental matter, $\nf \ll \nc$,  may be described by probe
D-branes in the appropriate gravitational background \cite{flavour}.
At sufficiently high temperature $T$, this contains a black hole \cite{witten}.
The purpose of this paper is to exhibit some universal features of
this system that only depend on these two facts. In particular, we
demonstrate the existence of a first order phase transition for the
fundamental matter, as follows:

At sufficiently small $T/\mq$  (where $\mq$ is the quark mass), the
brane tension is sufficient to overcome the attraction of the black
hole and hence the branes lie outside the horizon in a `Minkowski'
\begin{figure}[h]
\includegraphics[scale=0.8]{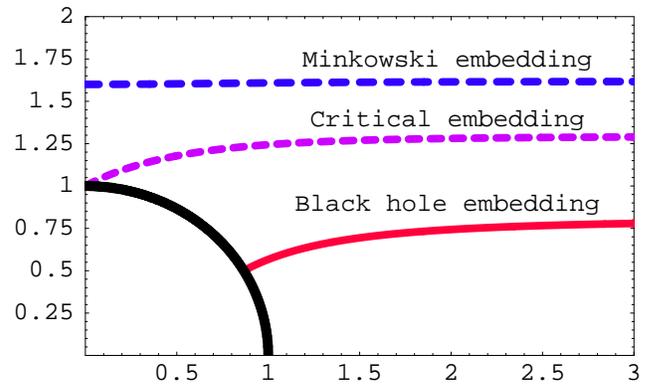}
\caption{Profiles of D7-brane embeddings in a D3-brane background.
The thick black circle is the horizon ($\rho =1$).}
\label{embeddings}
\end{figure}
embedding --- see fig.~\ref{embeddings} and below. In this phase the
meson spectrum (\ie the spectrum of quark-antiquark bound states) is
discrete and possesses a mass gap. At sufficiently large $T/\mq$,
the gravitational attraction overcomes the brane tension and the
branes fall into the horizon yielding a `black hole' embedding. In
this case, the meson spectrum is continuous and gapless. In between,
a limiting, critical solution exists. We will show that the phase
diagram in the vicinity of this solution exhibits a self-similar
structure. While this structure went unnoticed, the phase transition
that occurs as $T/\mq$ increases from small to large values was
observed in two specific models \cite{johanna,us}. In fact, as was
first noted in \cite{us} for a D6-brane in a thermal D4 background,
the transition is of first order (in the approximations stated
above). Rather than dropping continuously through the critical
solution, the probe brane jumps discontinuously from a Minkowski to
a black hole embedding at some $T=\tf$.  This leads to a
discontinuity in several field theory quantities, such as, for
example, the quark condensate $\qc$ or the entropy density. In the
following, we will see that the critical behavior and, as a result,
the first order transition are essentially universal to all Dp/Dq
systems.

The case of confining gauge theories is particularly interesting
because, for sufficiently heavy quarks, two distinct phase
transitions occur \cite{us}. At $T=\td$ the gravitational background
changes from a horizon-free background to a black hole background.
At this temperature the gluons and the adjoint matter become
deconfined  \cite{witten}, but the branes remain outside the horizon
and hence stable quark-antiquark bound states still exist in a range
$\td < T < \tf$. At $T=\tf$ the phase transition described above for
the fundamental matter occurs.
\newline
\noindent {\bf The Setup:} The black Dp-brane metric in the
string-frame takes the form \be ds^2 = H^{-\frac{1}{2}} \left( -f
dt^2 +  dx_{\it p}^2 \right) + H^{\frac{1}{2}} \left( \frac{du^2}{f}
+ u^2 d\Omega_{\it 8-p}^2 \right) \,, \label{metric} \ee where $H(u)
= (\R/u)^{7-p}$, $f(u) =1-(u_0/u)^{7-p}$ and $\R$ is a length scale
(the AdS radius in the case $p=3$). There is also a non-trivial
dilaton $e^\phi = g_s H^{(3-p)/4}$ and a Ramond-Ramond field  $C_{01\ldots p} =
H^{-1}$.  The horizon lies at $u=u_0$. As usual, regularity of the
Euclidean section, obtained through $t \rightarrow i\te$, requires
that $\te$ be identified with period
\be \frac{1}{T} = \frac{4\pi  \R}{7-p} \left( \frac{\R}{u_0}
\right)^{\frac{5-p}{2}} \,. \label{beta} \ee
According to the gauge/gravity correspondence, string theory on the
background above is dual to a $(p+1)$-dimensional supersymmetric
gauge theory at temperature $T$. In some cases one periodically
identifies some of the `Poincare' directions $x_{\it p}$ in order to
render the theory effectively lower-dimensional at low energies; a
prototypical example is that of a D4-brane with one compact space
direction. Under these circumstances a different background with no
black hole may describe the low-temperature physics, and a phase
transition may occur as $T$ increases \cite{witten}. In the gauge
theory this is typically a confinement/deconfinement phase
transition for the gluonic (or adjoint) degrees of freedom.
Throughout this paper we assume that $T$ is high enough, in which
case the appropriate gravitational background is always
\eqn{metric}.

Consider now a Dq-brane probe that shares $d$ spacelike `Poincar\'e'
directions with the background Dp-branes and wraps an $S^n$ inside
the $S^{8-p}$. We will assume that the Dq-brane also extends along
the radial direction, so that $q=d+n+1$. In the gauge theory this
corresponds to introducing fundamental matter that propagates along
a $(d+1)$-dimensional defect. To ensure stability, we will assume
that the Dp/Dq intersection under consideration is supersymmetric at
zero temperature. Under these conditions the RR field sourced by the
Dp-branes does not couple to the Dq-branes. Two cases of special
interest here are the D3/D7  ($n=3$) \cite{johanna} and the D4/D6
($n=2$) \cite{us} systems.  If one of the D4 directions is compact,
then both cases can effectively be thought of as describing the
dynamics of a four-dimensional gauge theory with fundamental matter.
\newline
\newline
\noindent {\bf Criticality and Scaling:} In this section we follow
\cite{frolovnew} closely. We begin by studying the behaviour of the
brane probe near the horizon. In order to do so we write \bea &&
d\Omega_{\it 8-p}^2 = d\theta^2 + \sin^2 \theta \, d\Omega_{\it n}^2
+ \cos^2 \theta \, d\Omega_{\it 7-p-n}^2  \,, \\ \label{coord} && u
= u_0 +  \pi T \Z^2 \,\,, \,\, \theta = \frac{\Y}{\R}  \left(
\frac{\R}{u_0} \right)^\frac{p-3}{4} \,\,, \,\, \tilde{x} = \left(
\frac{u_0}{\R} \right)^\frac{7-p}{4} x \nonumber \eea with $T$ the
temperature above. Expanding the metric to lowest order in $\Z, \Y $
gives Rindler space together with some spectator directions: \be
ds^2 = - (2\pi T)^2 \Z^2 dt^2 + d\Z^2 + d\Y^2  + \Y^2 d\Omega_{\it
n}^2 + d\tilde{x}_{\it d}^2 + \cdots \,. \ee The Dq-brane lies at
constant values of the omitted coordinates, so these play no role in
the following. The horizon is of course at $\Z=0$. The Dq-brane
embedding is specified by a curve $(\Z(\sigma), \Y(\sigma))$ in the
$(\Z,\Y)$-plane. Since the dilaton approaches a constant near the
horizon, up to an overall constant the Dq-brane (Euclidean) action
is simply the volume of the brane, namely \be I \propto \int d\sigma
\sqrt{\dot{\Z}^2 + \dot{\Y}^2} \, \Z \Y^n \,, \ee where the dot
denotes differentiation with respect to $\sigma$. This is precisely
the action considered in ref. \cite{frolovnew}. In the gauge
$\Z=\sigma$ the equation of motion takes the form \be \Z \Y
\ddot{\Y} + (\Y \dot{\Y} - n \Z) (1 + \dot{\Y}^2) = 0 \label{eom}
\,. \ee Solutions fall into two classes that we call  `black hole'
embeddings and `Minkowski' embeddings; see fig. \ref{embeddings}.
Black hole embeddings are those for which the brane falls into the
horizon, and may be characterised by the size of the induced
horizon, $\Y_0$. The appropriate boundary condition in this case is
$\dot{\Y}  = 0 , \Y = \Y_0$ at $\Z=0$. Minkowski embeddings are
those for which  the brane closes off smoothly above the horizon.
These are characterised by the distance to the horizon, $\Z_0$, and
satisfy the boundary condition $\dot{\Z} =0, \Z = \Z_0$ at $\Y=0$.
The limiting solution is the critical solution $\Y= \sqrt{n} \,\Z$,
which touches the horizon at the point $\Y=\Z=0$.

The equation of motion \eqn{eom} enjoys a scaling symmetry: If
$\Y=f(\Z)$ is a solution, then so is $\Y=f(\kk \Z)/ \kk$ for any
real positive $\kk$. This transformation rescales $\Z_0\rightarrow
\Z_0/ \kk$ for Minkowski embeddings, or $\Y_0 \rightarrow \Y_0/\kk$
for black hole embeddings, which implies that all solutions of a
given type can be generated from any other one by this
transformation.

Consider now a solution very close to the critical one, $\Y(\Z) =
\sqrt{n} \, \Z + \xi (\Z)$. Linearising the equation of motion
\eqn{eom}, one finds that for large $\Z$ the solutions are of the
form $\xi (\Z) = \Z^{\nu_\pm}$, with $\nu_\pm = -n/2 \pm \sqrt{n^2 -
4(n+1)} /2$. If $n \geq 5$ these exponents are real, whereas if
$n\leq 4$ they have non-vanishing imaginary parts that lead to
oscillatory behaviour. Under the assumptions stated above $n \geq 5$
implies $d \leq 1$ (an example of this is a D3/D7 intersection over
a string). Since we are mostly interested in higher dimensional
defects, we will  henceforth assume that $n\leq 4$. In this case it
is convenient to write the general solution as \be \Y =  \sqrt{n} \,
\Z + \frac{T^{-1}}{(T\Z)^{\frac{n}{2}}} \Big[ a \sin (\alpha \log T
\Z) + b \cos (\alpha \log T \Z) \Big] \,, \ee where
$\alpha=\sqrt{4(n+1)-n^2}/2$ and $a,b$ are dimensionless constants
determined by $\Z_0$ or $\Y_0$. It is easy to show that under the
rescaling discussed above, these constants transform as
\be
\begin{pmatrix}
a \cr b
\end{pmatrix}
\rightarrow \frac{1}{\kk^{\frac{n}{2}+1}}
\begin{pmatrix}
\cos (\alpha \log \kk) & \sin (\alpha \log \kk) \cr
-\sin (\alpha \log \kk) & \cos (\alpha \log \kk)
\end{pmatrix}
\begin{pmatrix}
  a\cr
  b
\end{pmatrix} \,.
\label{transf}
\ee
This transformation law implies that the solutions exhibit discrete
self-similarity and can be used to derive critical exponents that
characterise the near-critical behaviour. We refer the reader to
\cite{frolov,frolovnew} for details but we emphasise that this
behaviour depends only on the dimension of the sphere, and is
therefore universal for all Dp/Dq systems with $n\le4$.

Each near-horizon solution gives rise to a global solution when
extended over the full spacetime. Each of these solutions is
characterised by a quark mass $\mq$ and a quark condensate $\qc$
that can be read off from its asymptotic behaviour. Both of these
quantities are fixed by $\Z_0$ or $\Y_0$. As we will see, the values
corresponding to the critical solution, $\mqs$ and $\qcs$, give a
rough estimate of the point at which the phase transition occurs.
\newline
\newline
\noindent {\bf Fundamental Phase Transitions:} In order to study the
global solutions, it is convenient to introduce an isotropic,
dimensionless radial coordinate $\rho$ through
\be \left( u_0 \rho
\right)^{\frac{7-p}{2}} = u^{\frac{7-p}{2}} + \sqrt{ u^{7-p} -
u_0^{7-p}} \,. \ee
Note that the horizon is at $\rho=1$. Just for
concreteness, we will now assume that the Dp/Dq system under
consideration is T-dual to the D3/D7 one, in which case $(p-d) +(
n+1) =4$.
Under these circumstances, the Euclidean Dq-brane action density in the black Dp-brane
background is
\bea \frac{\ibk}{\N} = \int_{\rhomin}^\infty d\rho \left(
\frac{u}{u_0 \rho} \right)^{d-3} \left( 1 - \frac{1}{\rho^{2(7-p)}}
\right) \rho^n && \nn \times \, {(1-\chi^2)}^{\frac{n-1}{2}} \sqrt{
1-\chi^2 + \rho^2 \dot{\chi}^2  } \,, && \label{action} \eea
where $\N$ is a normalisation constant:
\be \N = \frac{\nf \tq  u_0^{n+1} \Omega_n}{4T} \,. \ee
$\tq=1/{(2\pi \ell_s)}^q g_s \ell_s$ is the Dq-brane tension,
$\Omega_n$ is the volume of a unit $n$-sphere, $\chi=\cos \theta$,
and \mbox{$\dot{\chi} =d\chi/d\rho$}. Multiple flavours arise by
introducing $\nf$ coincident probe branes. Up to a numerical
constant the normalisation factor is found to be
\be \N \sim \nf \nc T^d \leff(T)^{\frac{2(d-1)}{5-p}}  \,, \ee
where $\leff^2(T) = \lambda T^{p-3}$ is the dimensionless effective
`t Hooft coupling for a $(p+1)$-dimensional theory at temperature
$T$ \cite{itz}, $\lambda = \gym^2 \nc$ and we have used the standard
gauge/gravity relations $\gym^2 \sim g_s \ell_s^{p-3}$ and $\R^{7-p}
\sim g_s \nc \ell_s^{7-p}$ with $\ell_s$ the fundamental string
length.

The equation of motion that follows from \eqn{action} leads to the
large-$\rho$ behaviour\footnote{Here we assume $n>1$. For $n=1$, the
second term is  $c\,\log \rho/\rho$.}
\be \chi = \frac{m}{\rho} + \frac{c}{\rho^{n}} + \cdots \,.
\label{asymp} \ee
The constants $m,c$ are related to the quark mass and condensate
through \cite{johanna,us}
\be \mq = \frac{u_0 m}{2^{\frac{9-p}{7-p}}\pi \ell_s^2} \sac \qc = -
\frac{2 \pi \ell_s^2 (n-1) \Omega_n \tq u_0^n c} {4^{\frac{n}{7-p}}}
\,. \ee
This implies the relation $m^{(5-p)/2} = \mbar/T$ between the
dimensionless quantity $m$, the temperature $T$ and the mass scale
\be \bar{M} = \frac{7-p}{2^{\frac{9-p}{7-p}}\pi \R} \left(
\frac{2\pi \ell_s^2 \mq}{L} \right)^{\frac{5-p}{2}} \sim
\frac{\mq}{\leff(\mq)} \,, \label{deep} \ee
This is precisely the scale of the mass gap in the discrete meson
spectrum at temperatures well below the phase transition
\cite{us-meson,us,rob-rowan}. We see here that it is also the scale of the
transition temperature for the fundamental degrees of freedom, $\tf
\sim \mbar$, since this takes place at $m \sim 1$.

\begin{figure}[!]
\includegraphics[scale=.6]{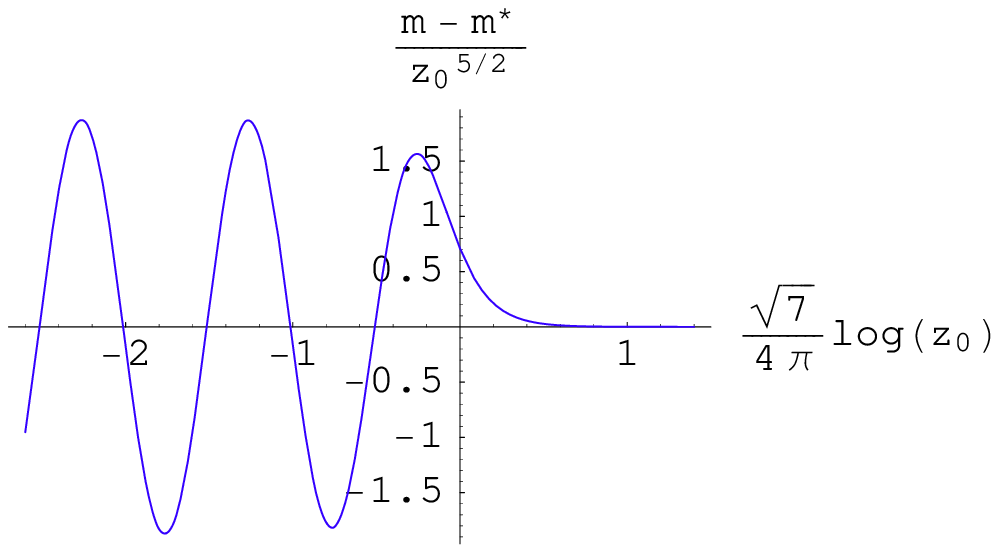}
\hspace{1mm}
\includegraphics[scale=.6]{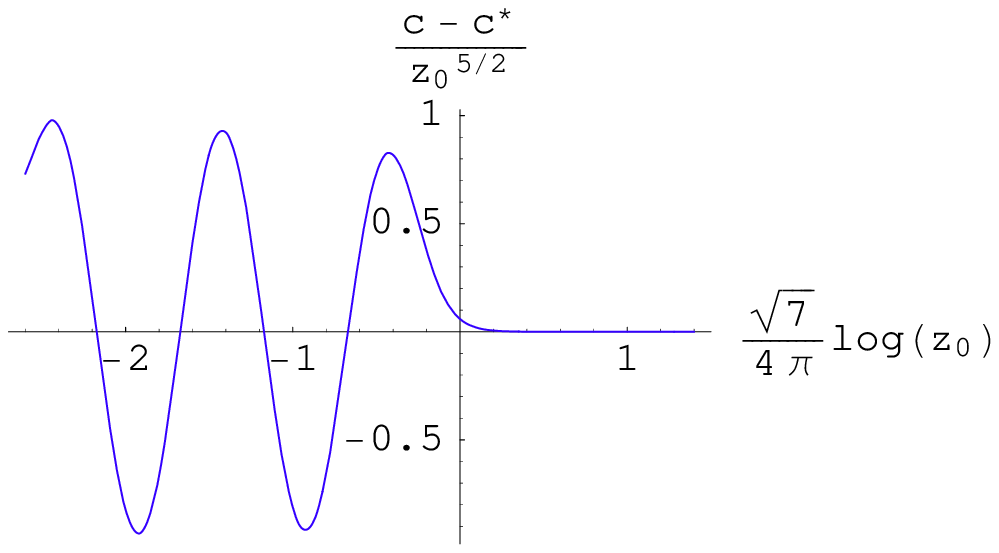}
\caption{Quark mass as a function of the distance to
the horizon $\Z_0$ for  D7-brane Minkowski embeddings in a D3-brane background.}
\label{mcfunc}
\end{figure}

The key observation \cite{frolov} is that the values $(m,c)$ of a
near-critical solution are linearly related to the corresponding
values in the near-horizon region. Combining this with the
transformation rule \eqn{transf} for the near-horizon constants
$(a,b)$ and eliminating $\kk$, we deduce that $(m-m^*) /
\Z_0^{\frac{n}{2}+1}$ and $(c-c^*) / \Z_0^{\frac{n}{2}+1}$ are
periodic functions of $(\alpha /2 \pi) \log \Z_0$ with unit period
for Minkowski embeddings, and similarly with $\Z_0$ replaced by
$\Y_0$ for black hole embeddings. This is confirmed by our numerical
results, as illustrated in fig. \ref{mcfunc}.

\begin{figure*}[!]
\includegraphics[scale=.6]{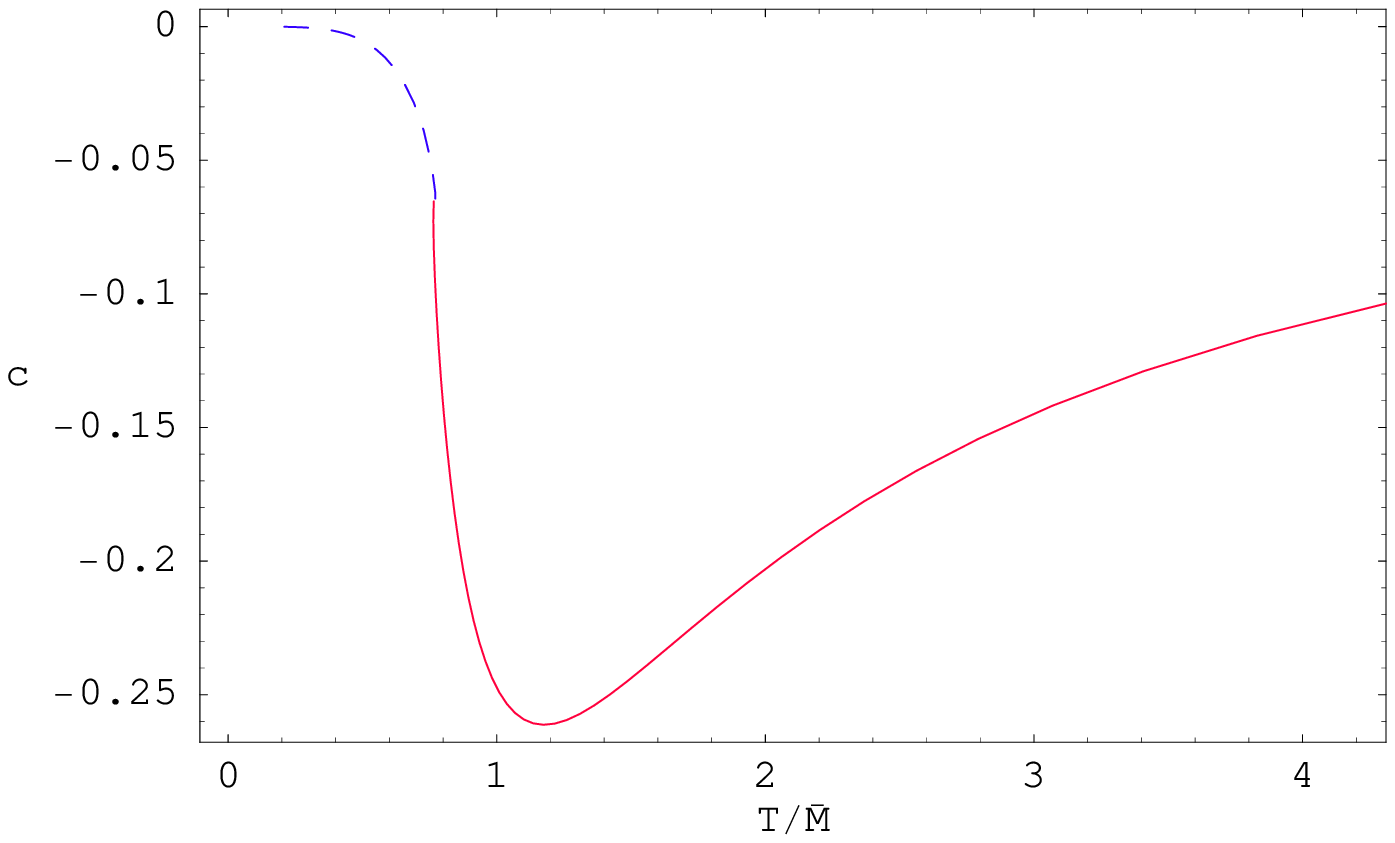}
\hspace{5mm}
\includegraphics[scale=.6]{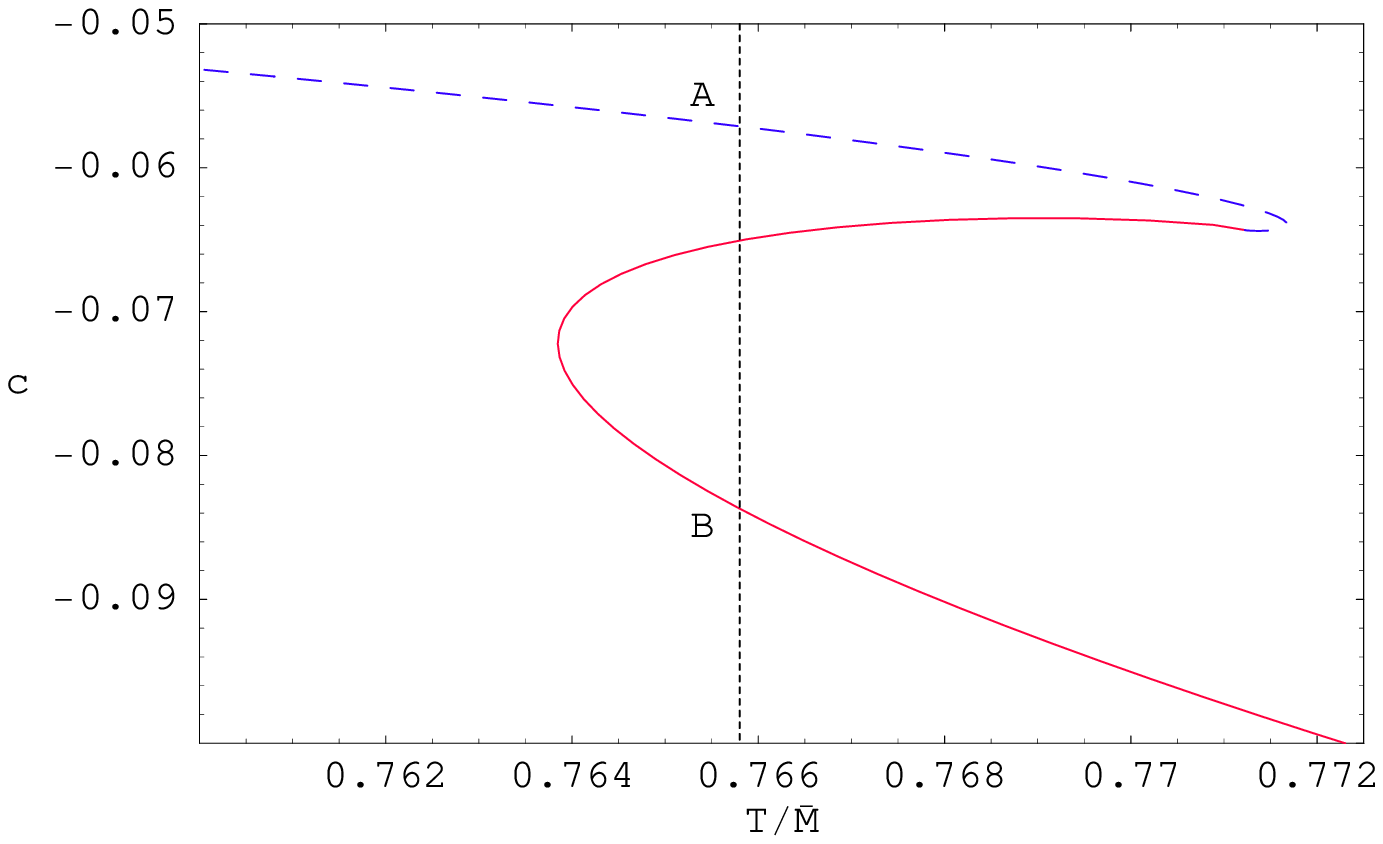} \\
\includegraphics[scale=.6]{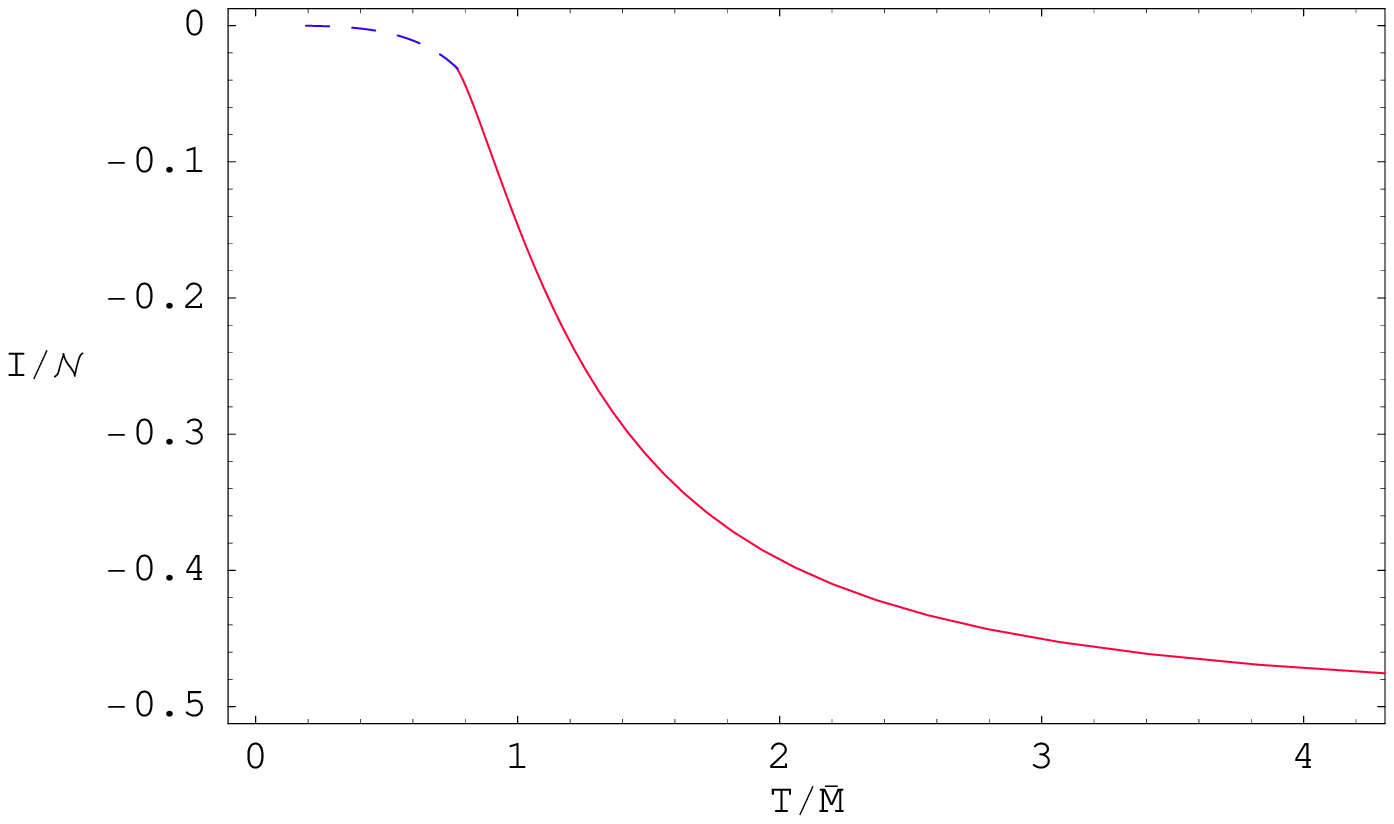}
\hspace{5mm}
\includegraphics[scale=.6]{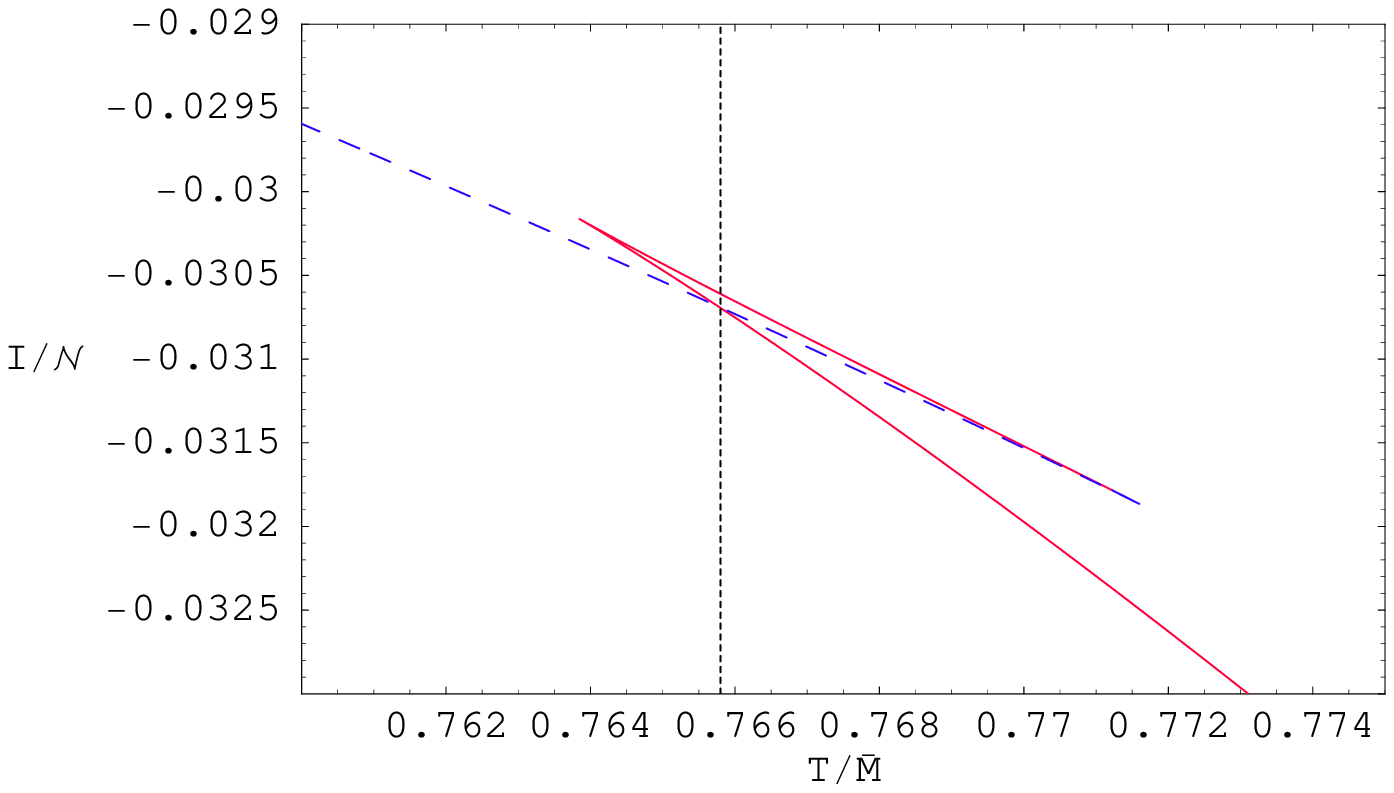}
\caption{Quark condensate  and free-energy density for a D7 in a D3 background;
note that $\N \propto T^3$. The blue dashed (red continuous) curves correspond
to the Minkowski (black hole) embeddings.  The dotted vertical line indicates
the precise temperature of the phase transition.} \label{bigfig}
\end{figure*}

The oscillatory behaviour of $m$ and $c$ as functions of $\Z_0$ or
$\Y_0$ implies that the quark condensate is not a single-valued
function of the quark mass. The preferred solution will of course be
the one that minimises the free energy density of the Dq-brane, $F=T
\ibk$. This quantity contains a volume divergence, as can be seen by
using the asymptotic behaviour \eqn{asymp}. It therefore needs to be
regularised and renormalised. We achieve the former by replacing the
upper limit of integration by a finite ultraviolet cut-off
$\rhomax$. The latter can be done by subtracting the free energy of
a fiducial embedding, as was done for the D4/D6 system in ref.
\cite{us}. The asymptotic geometry of a D7-brane in a D3 background
is $AdS_5 \times S^3$, so in this case the more elegant method of
holographic renormalisation \cite{ct} for brane probes \cite{karch1}
is available. This consists of adding to the brane action the
boundary `counter-term'
\be \frac{\ict}{\N} = - \frac{1}{4} \left((\rhomax^2 - m^2)^2 - 4 m
c  \right)  \,. \ee
The renormalised brane energy $I=\ibk + \ict$ is then finite as the
cut-off is removed, $\rhomax \rightarrow \infty$. The results for
the D3/D7 case are shown in fig. \ref{bigfig}. We see that the
transition occurs discontinuously by jumping from a Minkowski
embedding (point A) to a black hole embedding (point B). We
emphasize again that this first order transition is a direct
consequence of the multi-valued nature of the physical quantities
brought on by the critical behaviour described in the previous
section. It may be possible to access this self-similar region by
super-cooling the system.

It is interesting to ask if the strong coupling results obtained
here could in principle be compared with a weak coupling
calculation. It follows from our analysis that the free energy
density takes the form $F=\N T f(m^2)$, where the function $f$ can
only depend on even powers of $m$ because of the reflection symmetry
$\chi \rightarrow -\chi$. The strong coupling limit $\leff
\rightarrow \infty$ corresponds to $m \rightarrow 0$, which may be
equivalently regarded as a zero quark mass limit or as a
high-temperature limit. In this limit the brane lies on the
equatorial embedding $\chi=0$ and slices the horizon in two equal
parts. In general $f(0)$ is a non-zero numerical constant; in the
D3/D7 case, for example, a straightforward calculation yields
$f(0)=-1/2$. Thus  at strong coupling the free energy density scales
as
\be F \sim \nf \nc T^{d+1} \leff(T)^{\frac{2(d-1)}{5-p}}  \,.
\label{F} \ee
The temperature dependence is that expected on dimensional grounds
for a $d$-dimensional defect, and the $\nf \nc$ dependence follows
from large-$N$ counting rules. However, the dependence on the
effective 't Hooft coupling indicates that this contribution comes as a
strong coupling effect, without direct comparison to any weak
coupling result. The same is true for other thermodynamic quantities
such as, for example, the entropy density $S=-\partial F/\partial
T$. We remind the reader that the background geometry makes the
leading contribution to the free energy density \cite{polpee}
\be F \sim \nc^2 T^{p+1} \leff(T)^{\frac{2(p-3)}{5-p}} \,,
\label{F1} \ee
which corresponds to that coming from the gluons and adjoint matter.
Of course, for $p=3$, the effective coupling factor is absent and
this bulk contribution differs from the weak coupling result by only
a factor of 3/4 \cite{3/4}.
\newline
\noindent {\bf Discussion:} We have shown that, in a large class of
gauge theories with fundamental matter, quark-antiquark bound states
survive the deconfining phase transition for the gluonic degrees of
freedom provided $\mbar \gtrsim T_\mt{deconf}$, where $\mbar$ is the
typical mesonic scale. This is potentially interesting in connection
with QCD, since in QCD heavy quark mesonic bound states with $\mbar
\gg T_\mt{deconf} \sim 175$ MeV certainly exist. One generic feature
of the low-lying mesons in the class of theories discussed here is
that they are extremely deeply bound at strong coupling
\cite{us-meson,rob-rowan}, as is apparent from eq. \eqn{deep}. It is
intriguing that the mesonic states claimed to explain certain
features (such as the entropy density) of the strongly coupled
quark-gluon plasma formed at RHIC are also deeply bound
\cite{brown}. It would be remarkable if a precise relationship
between the two could be established.

Thermal phase transitions  in gauge theories with
spontaneously broken chiral symmetries at zero temperature
are particularly interesting, since they raise the question of whether
chiral symmetry is restored at the phase transition. This has been found
to be the case \cite{chiral} in the D4/D8/$\bar{\mbox{D8}}$ holographic
model of \cite{sugimoto}. In this model the D8/$\bar{\mbox{D8}}$ pair is
connected before it falls into the horizon, but splits into two
disconnected pieces when it does. At this point the
symmetry is enhanced from the diagonal $U(\nf)_\mt{V}$ subgroup
to the full $U(\nf)_\mt{L} \times U(\nf)_\mt{R}$ group. This contrasts
with the cases discussed in this paper, in which the branes remain
connected when they fall into the horizon.  From the gauge
theory viewpoint, this difference in topologies is due to the fact
that in the first case the fundamental matter lives on a defect/anti-defect pair,
whereas in the second it lives on a single defect.

The detailed results found in this paper rely on the approximation that
$1/\nc, 1/\lambda \rightarrow 0$ with $\nf$ fixed. However, the fact
that the phase transition is first order implies that it should be
stable under small perturbations, and so its order and other
qualitative details should hold within a finite radius of the $1/\nc,
1/\lambda$ expansions. Of course, finite-$\nc$ and finite-$\lambda$
corrections may
eventually modify the behaviour uncovered here. For example, at
large but finite $\nc$ the black hole will Hawking-radiate and each
bit of the brane probe will experience a thermal bath at a
temperature determined by the local acceleration. This effect
becomes more and more important as the lower part of a Minkowski
brane approaches the horizon, and may potentially blur the
self-similar, scaling behaviour found here. Note that this effect is
of order $1/\nc^2$, and therefore subleading with respect to the
order-$\nf/\nc$ correction to physical quantities from the presence
of the brane probes.

Finite 't Hooft coupling corrections correspond to higher-derivative
corrections both to the supergravity action and the D-brane action.
These may also blur the structure discussed above. For example,
higher-derivative corrections to the D-brane equation of motion are
likely to spoil the scaling symmetry of eq. \eqn{eom}, and hence the
self-similar behaviour. These corrections also become important as
the lower part of a Minkowski brane approaches the horizon, since
the (intrinsic) curvature of the brane becomes large there.

Yet another type of correction one may consider is due to the
backreaction on the background spacetime of the Dq-branes, whose
magnitude is controlled by the ratio $\nf/\nc$. These have been
considered for the D2/D6 system in ref. \cite{brandeis}. In
particular, this ref. finds that the energy density scales as $F
\sim \nf^{1/2} \nc^{3/2} T^3$, which obviously differs from \eqn{F}
with $p=2, d=2$. This discrepancy is not at all a contradiction, and
has the same origin as the discrepancy found for the meson spectrum
\cite{rob-rowan}. This is the fact that the calculation in
\cite{brandeis} applies in the far infrared of the gauge theory,
whereas that presented here applies at high temperatures, \ie at
$T\gg\gym^2$.

We will return to these and other issues in \cite{future}.

\noindent {\bf Acknowledgments:} We thank V. Frolov for a
conversation and for sharing ref.~\cite{frolovnew} with us prior to
publication.  We also thank P. Kovtun, A. Starinets and L. Yaffe for discussions. Research at the Perimeter
Institute is supported in part by funds from NSERC of Canada and
MEDT of Ontario. We also acknowledge support from NSF grant
PHY-0244764 (DM), NSERC Discovery grant (RCM) and NSERC Canadian
Graduate Scholarship (RMT). DM and RCM would also like to thank the
KITP for hospitality in the early stages of this project. Research
at the KITP was supported in part by the NSF under Grant No.
PHY99-07949.

\end{document}